\documentclass[vecphys]{svmult}

\usepackage{makeidx}         
\usepackage{graphicx}        
\usepackage{multicol}        
\usepackage[bottom]{footmisc}

\makeindex             

\begin{document}

\title*{The focal-plane instruments on board WSO-UV}
\author{Isabella Pagano\inst{1}\and
Mikhail Sachkov\inst{2}\and
Ana I. G\'omez de Castro\inst{3} \and
Maohai Huang\inst{4}\and
Norbert Kappelmann\inst{5}\and
Salvatore Scuderi\inst{1}\and
Boris Shustov\inst{2}\and
Klaus Werner\inst{5}\and
Gang Zhao\inst{4}}
\authorrunning{I. Pagano et al.}
\institute{INAF, Catania Astrophysical Observatory, Italy
\texttt{ipa,sscuderi@oact.inaf.it} \and
INASAN, Moscow, Russia \texttt{msachkov,bshustov@inasan.ru} \and
Fac. de CC. Matem\'aticas, Universidad Complutense de
Madrid, Spain \texttt{aig@mat.ucm.es}  \and
National Astronomical Observatories, CAS, China
\texttt{mhuang,gzhao@bao.ac.cn} \and
University of T\"ubingen, IAA, Germany
\texttt{kappel,werner@astro.uni-tuebingen.de}
}
%
%
\maketitle

\begin{abstract}
Dedicated to spectroscopic and imaging observations of the ultraviolet sky, the World Space Observatory for Ultraviolet Project is a Russia led international collaboration presently involving also China, Germany, Italy, Spain and Ukraine.
The mission consists of a 1.7m telescope able to perform: a) high resolution (R$\ge$60\,000) spectroscopy by means of two echelle spectrographs covering the 103-310 nm range; b) long slit (1$\times$75 arcsec) low resolution (R$\sim$1500-2500) spectroscopy using a near-UV channel and a far-UV channel to cover the 102-310nm range; c) deep UV and diffraction limited UV and optical imaging (from 115 to 700 nm).
Overall information on the project and on its science objectives are given by other two papers in these proceedings. Here we present the WSO-UV focal plane instruments, their status of implementation, and the expected performances.
\end{abstract}

\section{Introduction}
\label{sec:1}
The World Space Observatory-UV is an international collaboration led by Russia to build a space telescope dedicated mainly to UV astrophysics. The telescope is a Ritchey-Chretien with a diameter of 1.7 m, a F/10  focal ratio and a corrected field of view of 0.5 degrees. The primary wavelength range is 100-350 nm with an extension into the visible range. Specific data on the WSO-UV project (telescope, satellite, orbit, launcher, ground segment etc.) are given by Sachkov et al. (these proceedings). The WSO-UV telescope feeds in its focal plane a set of instruments for spectroscopy and imaging.

\begin{table}
\centering
\caption{Characteristics of the WSO-UV spectrographs}
\begin{tabular}{lcccc}
\hline
Spectrograph & Range & Spectral & Detector & Minimum\\
 & nm & Resolution & Sensitivity$^a$\\
\hline
 HIRDES-VUVES & 102--176 & $>$60\,000 & MCP & 16 mag\\
 HIRDES-UVES & 174--310 & $>$60,000 & MCP &18 mag\\
 LSS & 102--320 & 1\,500--2\,500 & MCP & \\
\hline
\multicolumn{5}{l}{SNR=10 in 10 h}\\
\end{tabular}
\label{tab:1}
\end{table}

\begin{description}

\item [{\bf HIRDES}:] The High Resolution Double Echelle Spectrograph (Funding Agency: DLR -- Principal Investigator: Prof. K. Werner, University of T\"ubingen, Germany -- Industrial Contractor: Kayser Threde), that delivers spectra at R$\ge$60\,000 in the 102-176 nm (VUVES), and in the 174-310 nm  (UVES) ranges.
\item [{\bf LSS}:] The Long Slit Spectrograph (Funding Agency: CNSA - Project Director: Prof. Gang Zhao, National Astronomical Observatories of Chinese Academy of Science), that delivers spectra in the range 102-320 nm, having  spectral resolution R$\sim$1500-2500, and allowing the use of a long slit of 1$\times$75 arcsec.
\item [{\bf FCU}:] The Field Camera Unit (Funding Agency: ASI -- Principal Investigator: Dr. I. Pagano, INAF, Catania Astrophysical Observatory -- Industrial contractors: Galileo Avionica and Thales Alenia Space IT-MI), that provides images, low resolution field spectroscopy, and polarimetry in three channels: FUV (115-190 nm), NUV (150-280 nm), and UVO (200-700 nm) with spatial resolution of 0.2, 0.03, and 0.07 arcsec pixel$^{-1}$ and field of view of 6.6$\times$6.6, 1$\times$1, and 4.7$\times$4.7 arcmin$^{2}$, respectively.
\end{description}

In the next sections we describe the characteristics and performances of these science instruments.

\begin{table}
\centering
\caption{Characteristics of the FCU imager}
\begin{tabular}{lccccccc}
\hline
Channel & Range & Spatial & FoV &Detector & Array & Limiting\\
 &  & Resolution$^a$ &  &  & Size&  Flux$^b$\\
  & $nm$ &$arcsec$  & $arcmin^2$& & & erg\,cm$^{-2}$\,s$^{-1}$\,\AA$^{-1}$\\
\hline
 FUV & 115--190 & 0.2~ & 6.6$\times$6.6 & MCP(CsI) & 2k$\times$2k & 5.5$\times$10$^{-16}$\\
 NUV & 150--280 & 0.06 & 1$\times$1 & MCP(CsTe) &2k$\times$2k & 5.0$\times$10$^{-16}$\\
 UVO & 200--700 & 0.07 &  4.7$\times$4.7 & CCD$^c$ & 4k$\times$4k & 1.8$\times$10$^{-16}$\\
\hline
\multicolumn{8}{l}{a) per pixel element}\\
\multicolumn{8}{l}{b) SNR=10 in 1 h @[150, 255, 550]nm for [FUV,NUV,UVO], respectively}\\
\multicolumn{8}{l}{c) UV Optimized}\\
\end{tabular}
\label{tab:2}
\end{table}

\section{The instrument compartment}
\begin{figure}
\centering
\includegraphics[height=6cm]{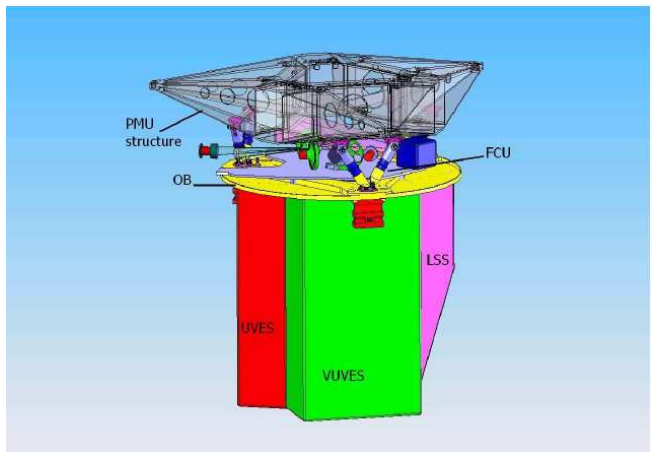}
%
%
\caption{The WSO-UV instrument compartment.}
\label{fig:SIC}       
\end{figure}

The WSO-UV telescope provides an accessible field of view of 30 arcmin on the telescope focal plane. In the instrument compartment, see Figure~\ref{fig:SIC}, the optical bench (OB)  -- used as reference plane for all the onboard instrumentation -- is aligned and maintained in the correct position with respect to the primary mirror unit (PMU) using a three rods system. The FCU is mounted on the upper basis of the optical bench, in the space available between the PMU and the OB itself, while the three spectrographs UVES, VUVES and LSS are mounted to the OB bottom basis. The Fine Guidance System, that uses three visible sensors placed at the vertex of an equilater triangle in the focal plane close to the spectrograph entrance slits, ensures a pointing stability better than 0.03 arcsec (1$\sigma$).

Each of the focal science plane instruments has its own power supply and data handling unit in a service box mounted on the external side of the instrument compartment.

\section{The High Resolution Double Echelle Spectrograph: UVES and VUVES channels}
\begin{figure}
\centering
\includegraphics[height=6cm]{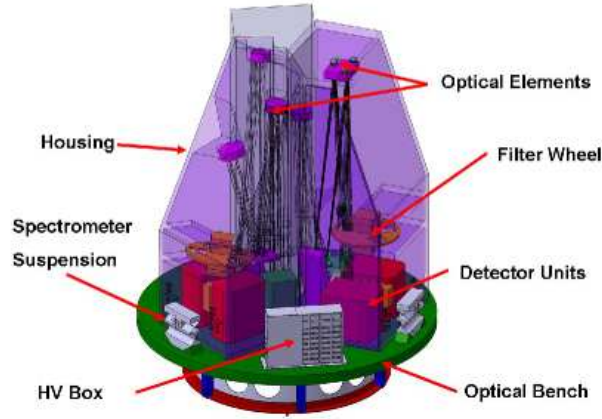}
%
%
\caption{HIRDES design at the end of the Phase B1 study.}
\label{fig:HIRDES}       
\end{figure}

The High Resolution Double Echelle Spectrograph (HIRDES) is made by two echelle
instruments, VUVES (102-176 nm), and UVES (174-310 nm) able to deliver high  resolution spectra (R$\ge$60,000) (see Figure~\ref{fig:HIRDES}). The HIRDES design uses the heritage of the ORFEUS ({\em Orbiting and
Retrievable Far and Extreme Ultraviolet Spectrometer}) missions
successfully flown on the Space Shuttle in 1993 and 1996 \cite{Barn99}.
The entrance slits of the two spectrographs lie in the focal
plane, on a circle with diameter 100 mm
which also hosts the LSS slits. The two channels can be operated in
sequential mode, by moving the satellite with a pointing
stability of 0.1 arcsec.
The position of the target in the slit is monitored by a  (visible) sensor of  an Internal Fine Guidance System, which is part of every  spectrograph.
 The VUVES and UVES detectors
are photon counting devices based on Microchannel Plates, readout by
means of a Wedge\&Strip Anode based on the ORFEUS detector design.
The main characteristics of the HIRDES spectrographs are listed in Table~\ref{tab:1}

 A Phase B1 study for
VUVES and UVES has been completed in May 2006 by Germany, with
Russia collaboration and with industrial support by Kayser--Threde.

The limiting magnitudes (SNR=10 in 10h) are
18 and 16 for UVES and VUVES, respectively. An Exposure Time
calculator is available on the web
(http://astro.uni-tuebingen.de/groups/wso\_uv/exptime\_calc.shtml).

\begin{figure}
\centering
\includegraphics[height=4.35cm]{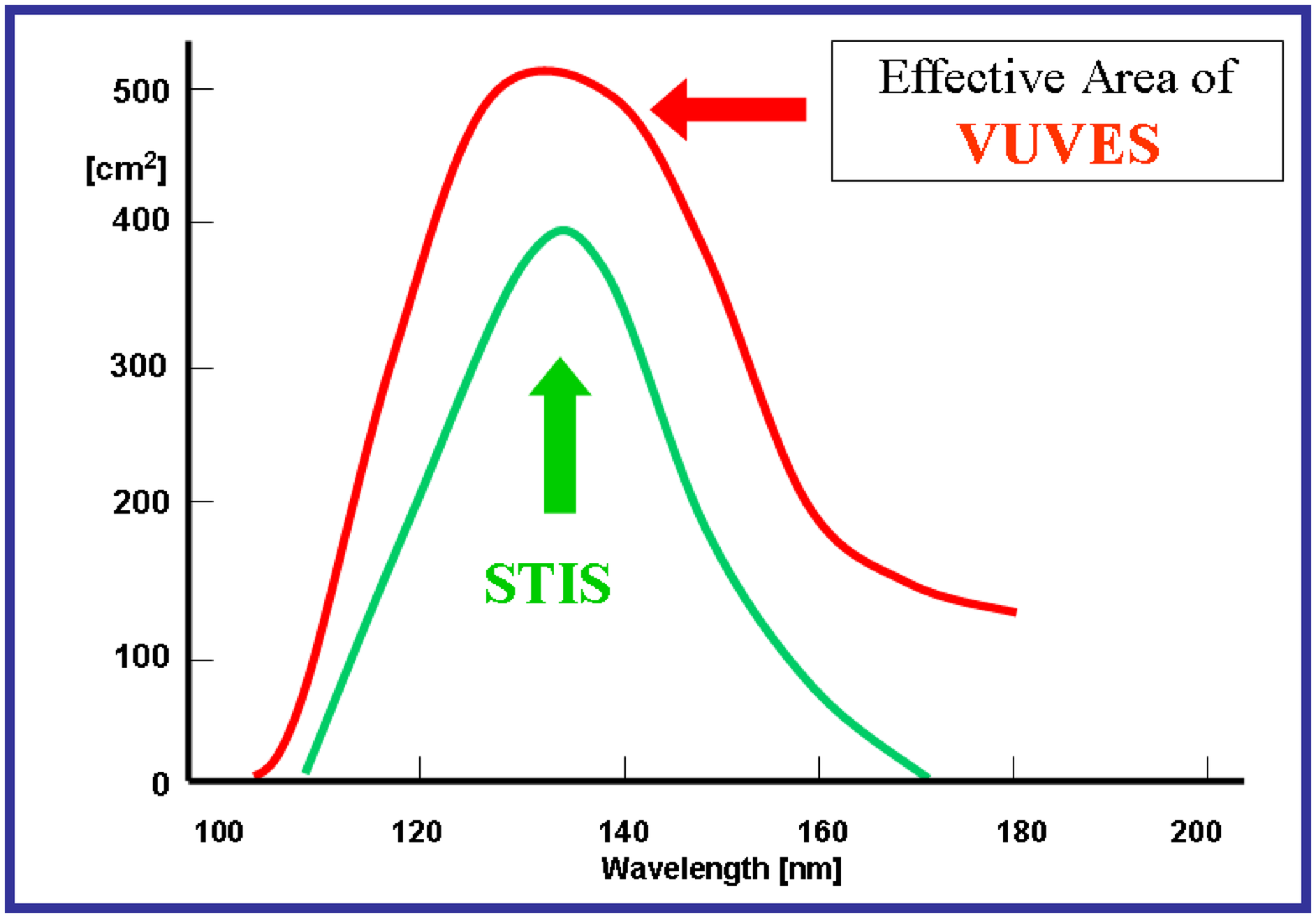}
\includegraphics[height=4.35cm]{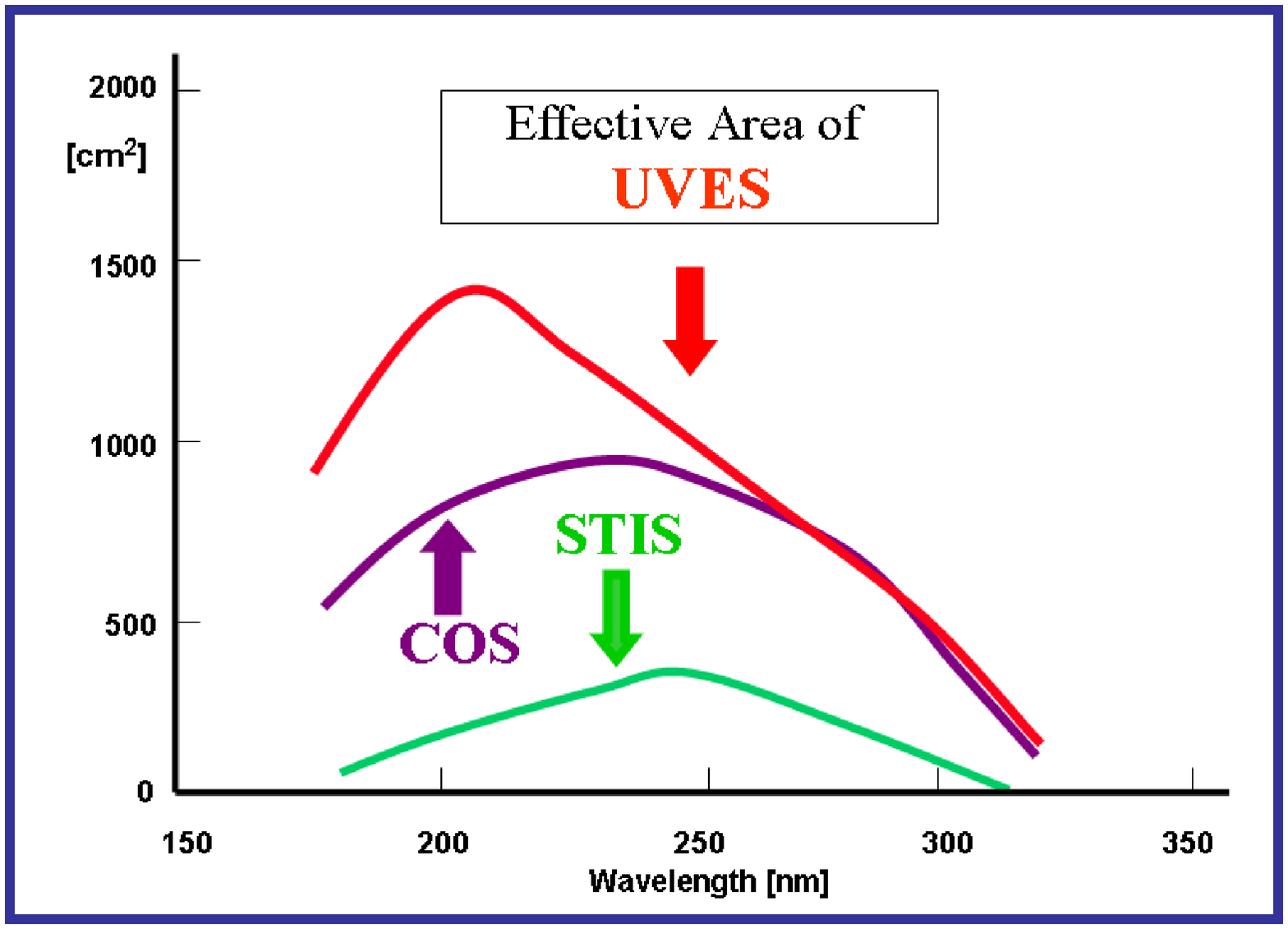}
%
%
\caption{HIRDES VUVES \& UVES effective area vs. wavelength compared to that of
the HST STIS and COS spectrographs. The VUVES channel has an effective area a factor $\sim$ 4 less than HST/COS.}
\label{HIRDES-effa}       
\end{figure}

The spectral resolution provided by HIRDES is similar to that
offered by HST/STIS at medium resolution with its echelle gratings, but higher than the maximum resolution provided by HST/COS (R$\sim$20\,000). The VUVES channel has an effective area a factor $\sim$ 4 less than HST/COS, while -- as shown in Figure~\ref{HIRDES-effa} -- the UVES channel effective area  is from 2 times more to the same one of HST/COS, as function of wavelength. In comparison with STIS, the HIRDES effective area is higher at every wavelength.

\section{The Long Slit Spectrograph}

The Long Slit Spectrograph will provide low resolution
(R$\sim$1500-2500) spectra in the 102-320 nm spectral range with a design that emphasizes sensitivity for observing faint objects.

\begin{figure}
\centering
\includegraphics[height=10cm,angle=270]{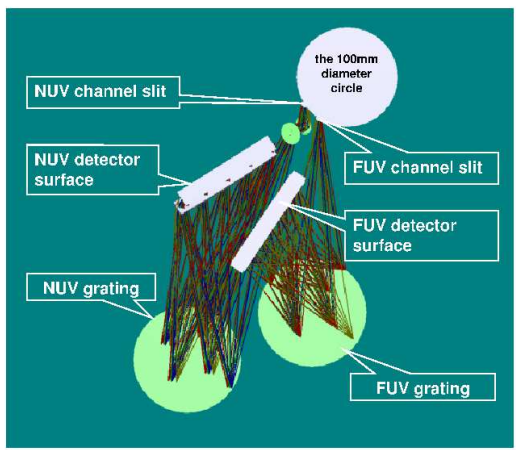}
%
%
\caption{LSS viewed from the direction of telescope with
 an oblique angle to show schematic layout of main optical
 and detector elements in LSS Phase A study. The
 100mm circle is on the focal plane around the optical axis of
 the primary mirror, and is where
 HIRDES and LSS entrance slits are placed (See text).}
\label{fig:LSS}       
\end{figure}

According to preliminary results of phase A study on-going at NAOC (China),
the most promising design for this instrument is to have two
channels working in 102-161 nm (FUV) and
160-320 nm (NUV) ranges, respectively, as shown in Figure~\ref{fig:LSS}.
Each channel has its own entrance slit, the arrangement of which is
 described in the HIRDES instrument section above. Both slits are
1"$\times$75" wide that allows 1 arcsec  spatial resolution.

In order
to maximize sensitivity both LSS channels use holographic gratings to
minimize the number of reflecting surfaces in optical path, and to
enhance overall spatial resolution along the slit. Both channels use
microchannel plates working in photon-counting modes as detectors.
A slit-viewer similar to that is used in HIRDES, internal calibrator,
and detector redundancy are being studied.

The two LSS channels can be operated independently. To obtain spectra
of full spectral range of the same source the telescope is required
to re-point when switching channel. Besides normal integration mode,
special observing modes including high time-resolution mode
(up to 1 second) and ``windowed mode'', whereas only photons
detected by specific part of the detector
surface are recorded, are envisaged.

\section{The Field Camera Unit}
The highest priorities of the FCU imager is to guarantee high spatial resolution and high UV sensitivity while trying to maximize the wavelength coverage and the size of the field of view. The science requirements for the the imagers in the focal plane of WSO-UV are:
\begin{itemize}
\item the possibility to obtain diffraction limited images for $\lambda >$230 nm (goal $\lambda >$200 nm).
\item S/N=10 per pixel in one hour exposure time down to:
 \begin{itemize}
\item V=26.2 (21.5--11.1) for a O3 V (A0 V--G0 V) star at 150 nm
\item V=24.6 (21.9--19.5) for a O3 V (A0 V--G0 V) star at 250 nm
\item V=26.3 (26.3--26.3) for a O3 V (A0 V--G0 V) star at 555 nm
\end{itemize}
\item Spectral resolution R=100 at $\lambda$=150 nm, R$\ge$100 at $\lambda$=250 nm, and R=250 at $\lambda$=500 nm.
\item Polarimetric filters
\item High time resolution imaging
\item The possibility to obtain wide field ($\sim$5$\times$5 arcmin$^2$) images from Far-UV to visual wavelengths.
\item Partial overlap between spectral range covered by FUV and NUV and NUV and UVO, for relative calibration purposes.
\end{itemize}

\begin{figure}
\centering
\includegraphics[height=6cm]{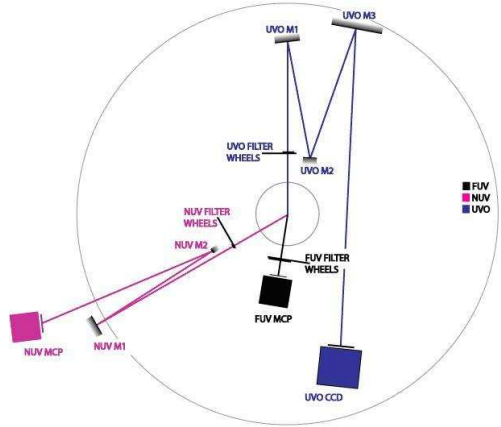}
%
%
\caption{FCU schematic design.}
\label{fig:FCU}       
\end{figure}

The instrument design concept resulting from the above requirements
is illustrated in Figure~\ref{fig:FCU}. Three wide band channels (FUV, NUV and UVO) are provided, covering a wavelength range from 115 nm to at least 700 nm (see Table~\ref{tab:2}). Each channel is specialized in a specific wavelength range and will have also spectroscopic, polarimetric and spectropolarimetric capabilities.

The Far-UV channel will cover the 115-190 nm range. To reduce losses in the throughput, this channel will not have any optics but the mirror to feed it. The scale of the telescope will be unchanged yielding a large field of view at expenses of spatial resolution. The FUV channel filter wheels will accommodate narrow and wide band filters and a disperser which will allow low resolution (R$\sim$100) slitless spectroscopy. The FUV channel will use a photon counting detector based on a MCP detector with a 2k$\times$2k format. It will have a CsI photocathode directly deposited on the MCP. The read-out system will be based on a CCD detector.

The Near-UV channel will cover the 150-280 nm range overlapping the FUV range on the shorter wavelengths and the UVO range on the longer ones. To exploit the diffraction limited optical quality of the telescope in this wavelength range the NUV channel has the highest spatial resolution. Its filter wheels will accommodate wide and narrow band filters and polarizers for imaging and imaging polarimetry. A grating will allow low resolution (R$\ge$100) slitless imaging spectroscopy. A spectro-polarimetric observational mode is under investigation. The NUV detector will differ from the FUV detector only for the photocathode ($Cs_2Te$), optimized for this wavelength range and deposited on the detector window.
\begin{figure}
\centering
\includegraphics[height=8cm]{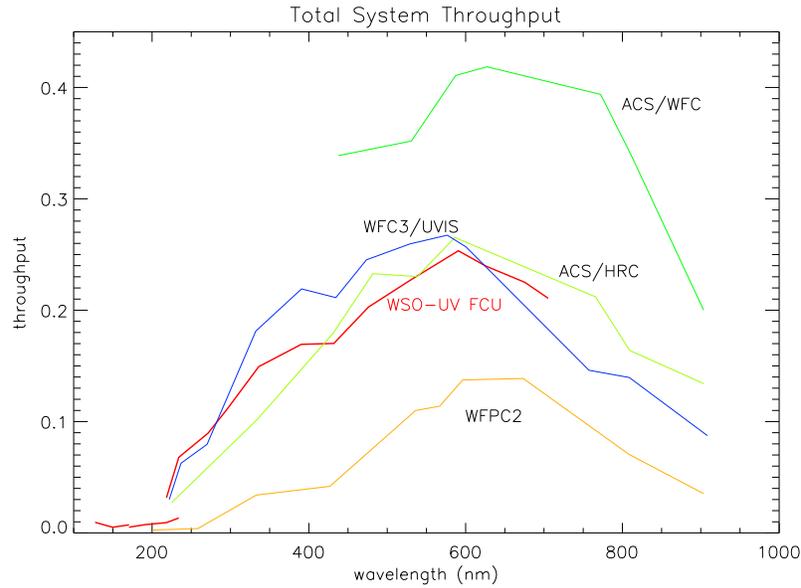}
%
%
\caption{System throughputs of the FCU vs. wavelength compared to that of
the HST imaging instruments.  Note
that the y-axis is logarithmic.
}
\label{throughput}       
\end{figure}
Finally the UVO channel will extend to visual wavelengths to exploit the wide spectral sensitivity of its CCD detector. The pixel scale of this channel is a compromise between the need of a large field of view and high spatial resolution. Filters, dispersers and polarizers will allow to have narrow and broad band imaging, low resolution (R$\sim$100-500) slitless spectroscopy and imaging polarimetry. The UVO channel will use a single 4k$\times$4k CCD detector optimized for the UV.

The geometry of the focal plane of the telescope makes it necessary to use a mirror to fold the optical beam coming from the telescope in a direction parallel to the optical bench where the three channels of the FCU will be deployed.
Two alternative optical layouts are investigated:
\begin{enumerate}
\item Rotating pick up mirror: in this case the mirror will be flat and -- using a mechanism -- will rotate feeding one of the three channels at a time.
\item Mosaic pick mirror: this layout foresees to use three fixed mirrors which will be oriented to feed all channels at the same time.
\end{enumerate}
The trade-off among the two solutions will be done during the phase B1 study, that will be completed by the end of 2007.

\begin{figure}
\centering
\includegraphics[height=8cm]{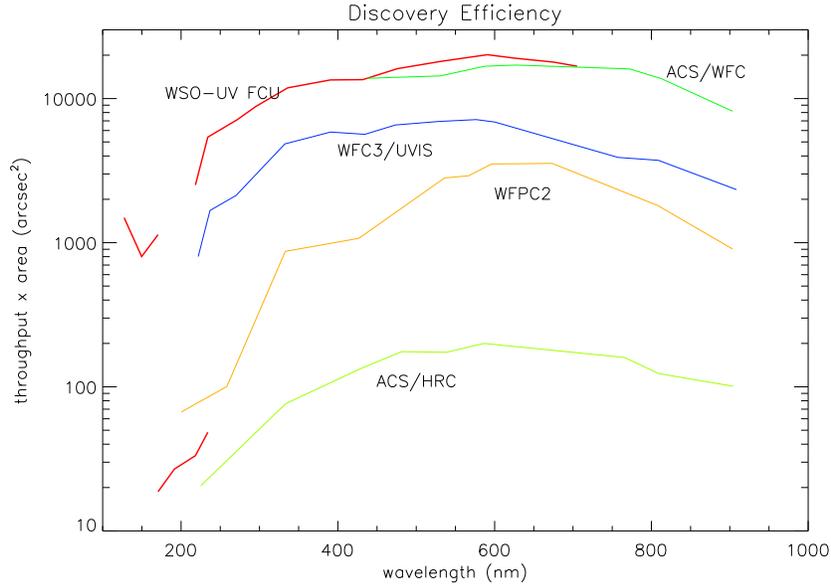}
%
%
\caption{System discovery efficiencies of the FCU vs. wavelength compared to that of
the HST imaging instruments. Discovery efficiency is defined as the
system throughput multiplied by the area of the field of view. Note
that the y-axis is logarithmic.
}
\label{de}       
\end{figure}
The expected system throughputs are compared in Figure~\ref{throughput}  to the ones of the cameras that have
flown or will fly on board the Hubble Space Telescope i.e. WFPC2,
ACS/WFC, ACS/HRC and WFC3/UVIS \cite{bond}. Throughput includes the telescope, all of the optical
elements of the instruments themselves, the sensitivities of the
detectors and filter transmissions, that for the FCU are estimated
based on the best information currently available and are subject
to change.
The throughput of the UVO channel is much better than that of WFPC2
which, by the way, is the oldest of the HST cameras, and is comparable
to that of ACS/HRC and WFC3/UVIS being better in the UV range.
ACS/WFC, being optimized in the visual range, has a much better
throughput than the UVO channel.

The discovery
efficiency, defined as system throughput times the area of the FOV as
projected onto the sky, is useful when comparing different instruments in the context of wide-angle surveys. Due to its large field of view the
UVO channel of the FCU has a discovery efficiency equal or greater than that of
ACS/WFC, as shown in Figure~\ref{de}. The performance of the FUV channel is even better when compared
to HST because, in this case (by the way ACS/SBC is not present in the
plot), no camera working in this wavelength range has a large field of view.

\subsection{Acknowledgments}
IPa and SSc thank colleagues of the Italian WSO-UV Team. The participation in the WSO-UV project in Italy is funded by Italian Space Agency under contract ASI/INAF No. I/085/06/0. AIGdC thanks the support by the Ministry of Science and Education of Spain
through grant ESP2006-27265-E. MH thanks colleagues of Chinese WSO-UV team. The HIRDES B1 Study in Germany was funded by the Federal Ministry of Education and Research by the German Aerospace Center (DLR) under contract No 50 QV 0503.


\printindex

\begin{thebibliography}{99.}
%
%
%




\bibitem{Barn99} Barnstedt J. et al.: A\&ASS, \textbf{134}, 561 (1999)

\bibitem{bond}
Bond, H.E., et al.: Wide Field Camera 3 Instrument
Mini-Handbook, Version 3.0, Baltimore: STScI (2006)


\end{thebibliography}
\end{document}